# On the quantum tunneling of one dimensional Coulomb singular potential barrier

Atom Muradyan and Gevorg Muradyan

Laboratory for research and modeling of quantum phenomena, Yerevan State University, 0025, 1 Alex Manoogian, Yerevan, Republic of Armenia

**E-mail: muradyan@ysu.am, gmurad@ysu.am**

**Abstract**

To put the quantum tunneling problem of a singular potential on a physical basis, two methods are usually used: the limiting cut-off procedure of the region of singularity and the matching of the wave function and its first derivative at two sides of the singular point. These approaches, nevertheless, effectively suppress the mathematical essence of the singularity. Hence the natural question of how quantum tunneling will behave when the singularity is preserved as much as possible is the main question of this paper. We get that the Coulomb singularity is reflected as infinitely accelerating oscillations in the transmission coefficient between zero and one when the incident particle's energy approaches the zero boundary. At relatively high energies, the tunneling acquires the character inherent in regular potentials, and becomes completely transparent in the asymptote.

## 1. Introduction

The principal problem in the study of quantum tunneling of singular potentials is not so much the infinite growing of the potential energy and its derivative when approaching the singular point, but the fact that this point is outside the domain of the definition of the potential energy function [1, 2]. At the same time, quantum tunneling requires a transition through the singular point and therefore some rules of this transition. To meet the physics of the problem, a potential cutoff method is used, or matching conditions are introduced for the wave function and its derivative. The first approach involves replacing the singular form with a regular one with the singular part cut off, calculating the transmission and reflection coefficients, and limit transition that narrows the truncation width to zero. The second approach proceeds from a self-adjoint extension of the Hamiltonian operator to the Hilbert space, for which applicability to singular potentials is limited by the Friedrichs extension [3]. These additions ensure the reality of the eigenvalues of the Hamilton operator defined on the entire coordinate axis, and thereby are a sufficient condition for handling any quantum mechanical problem. Simultaneously, it is natural that they effectively suppress the essence of the singular definition of the potential by turning it into a continuous potential with a certain (including infinite) value at any point. This statement can be justified in parallel by the fact that in the theory of differential equations, the sequence of two operations (such as calculating a certain value and a limit transition) is interchangeable only for finite continuous transformations. The singular nature is beyond these limits.

The subject of interet, quantum tunneling of a one-dimensional Coulomb potential barrier is, of course, a studied problem. M. Andrews in [1], by classifying the Coulomb potential as "singular", preserved only regular of the two linearly independent solutions. This automatically resulted in a lack of probabilistic flow and, consequently, impenetrability of the potential barrier. The same result was obtained by the method of limiting smoothing of the potential barrier and led to the conclusion that under a physically acceptable interpretation of the singularity, the one-dimensional Coulomb potential is impenetrable. The conclusion was then supported by C. Hammer and T. Weber during the presentation of a short letter [4]. Further, M. Moshinsky, considering unbounded states of an antisymmetric one-dimensional Coulomb potential [5], considers it preferable not to discard an irregular solution, but to transform it by a certain procedure into a regular one. Then the problem gets two linearly independent regular solutions and, accordingly, finite permeability for the Coulomb potential. R. Newton in [6] found this approach ingenious, but not sufficiently justified, in fact requiring remaining within such matching conditions that directly follow from the Schrodinger equation. In his response [7], M. Moshinsky substantiated a viewpoint that the criticism of R. Newton is relevant to the symmetric, but not to the antisymmetric form of the potential, which was considered by him.

V. Mineev [8] approached the problem from the point of view of general analytical continuation of even solution through the singularity point. In this case, any permeability other than full becomes possible: we only need to select the appropriate type of self-adjoint expansion. C. de Oliveira and A. Verri in [9] reexamined the permeability problem on base of self-adjoint extensions and explicitly defined conditions, when the origin is permeable or impermeable for the probability current density. It is also emphasized that the important Dirichlet boundary condition implies that the origin is impermeable. Finally, in [10] G. Abramovici and Y. Avishai developed an approach to the problem (by the way, it can be regarded as an alternative to Moshinsky's approach in [5]), which cancels the singularity in the appropriate bilinear form of the wave function. They arrived to the conclusion of a complete reflection from the potential.

Our goal in this paper is to consider the problem of how quantum tunneling of a one-dimensional Coulomb potential behaves while preserving, as far as possible, the singular definition of the potential. We refer only to the flow of a continuous wave function, without applying regularization or self-adjoint decomposition methods. It is found that the singular essence of the potential function manifests itself in infinitely accelerating oscillations with an approach to the zero edge of particle energy. In the high-energy asymptote, the transmission coefficient tends to one, as for any regular potential barrier.

## 2. Mathematical model and barrier transmission

The one-dimensional Schrodinger equation for a stationary state reads

$$\frac{d^2 \psi(z)}{d z^2} + \left(\varepsilon - \frac{u_0}{|z|}\right)\psi(z) = 0, \qquad (1)$$

where the space coordinate $z$ is normalized to an arbitrary length $l > 0$, $\varepsilon \geq 0$ and $u_0 > 0$ are the energy and "power" of the Coulomb potential, respectively, normalized to the "recoil" energy $E_{rec} = \hbar^2 / 2ml^2$.

On the right side of the origin ($z > 0$), linearly independent solutions of equation (1) are given by continuously differentiable expressions

$$\psi_{r,1}(z) = e^{-i\sqrt{\varepsilon}z} z \, {}_1F_1\left(1 - \frac{iu_0}{2\sqrt{\varepsilon}}, 2, i2\sqrt{\varepsilon}\, z\right), \qquad (2a)$$

$$\psi_{r,2}(z) = e^{-i\sqrt{\varepsilon}z} z \, U\left(1 - \frac{iu_0}{2\sqrt{\varepsilon}}, 2, i2\sqrt{\varepsilon}\, z\right), \qquad (2b)$$

where ${}_1F_1(\cdots)$ and $U(\cdots)$ are the hypergeometric functions of Kummer and Tricomi, respectively. The first solution is regular: the function itself and its derivative have a finite limit when approaching the singular point. The singular content of the potential function is as fully manifested in the second, irregular solution in the form of a logarithmic divergence of its derivative.

The general solution is written as

$$\psi_r(z) = a_{r,1}\psi_{r,1}(z) + a_{r,2}\psi_{r,2}(z) \qquad (3)$$

For the left side of the origin ($z < 0$), we have two equivalent choices of linearly independent solutions that are consonant with (2a) and (2b). They are obtained by substituting $u_0 \to -u_0$ or $z \to -z$, respectively.
We adopt the second one and write down accordingly

$$\psi_{l,1}(z) = -e^{i\sqrt{\varepsilon}z} z\,_1F_1\left(1 - \frac{iu_0}{2\sqrt{\varepsilon}}, 2, -i2\sqrt{\varepsilon}\,z\right), \tag{4a}$$

$$\psi_{l,2}(z) = -e^{i\sqrt{\varepsilon}z} z\,U\left(1 - \frac{iu_0}{2\sqrt{\varepsilon}}, 2, -i2\sqrt{\varepsilon}\,z\right), \tag{4b}$$

as linearly independent solutions and

$$\psi_l(z) = a_{l,1}\psi_{l,1}(z) + a_{l,2}\psi_{l,2}(z) \tag{5}$$

for a general solutions.

We consider the limit values of the wave function on both sides of the singular point equal, which directly implies the equality of the coefficients before the irregular solutions in (3) and (5):

$$a_{l,2} = a_{r,2}. \tag{6}$$

Note that this does not affect the presence of a singularity in general solutions, but assumes that it behaves in the same manner on both sides of the origin. Next, we turn to the quantity that follows directly from the problem statement, the probability flow. Substituting expressions (5) and (3) in the current definition

$$j = i\kappa \int (\psi\psi^{*\prime} - \psi^*\psi')dz \tag{7}$$

where $\kappa = \hbar/2ml$ is a constant and the derivative is with respect to the dimensionless $z$, we get

$$j_l = |a_{l,1}|^2 j_{l,11} + a_{l,2}^* a_{l,1} j_{l,12} + a_{l,1}^* a_{l,2} j_{l,21} + |a_{l,2}|^2 j_{l,22}, \tag{8a}$$

$$j_r = |a_{r,1}|^2 j_{r,11} + a_{r,2}^* a_{r,1} j_{r,12} + a_{r,1}^* a_{r,2} j_{r,21} + |a_{r,2}|^2 j_{r,22}, \tag{8b}$$

to the left and right of the singular point, respectively, where

$$j_{l,mn} = i\kappa \int (\psi_{l,m}\psi_{l,n}^{\prime *} - \psi_{l,m}'\psi_{l,n}^*)dz, \quad j_{r,mn} = i\kappa \int (\psi_{r,m}\psi_{r,n}^{\prime *} - \psi_{r,m}'\psi_{r,n}^*)dz, \quad m,n = 1,2$$

are the composite parts of the probability currents.

Since $\psi_{l,1}(z)$ and $\psi_{r,1}(z)$ are real functions, $j_{l,11}$ and $j_{r,11}$ in (7a) and (7b) respectively, identically equal to zero. In addition, $j_{l,21} = j_{l,12}^*$, $j_{r,21} = j_{r,12}^*$ by definition and $j_{l,12} = -j_{r,12}$, $j_{l,22} = -j_{r,22}$. Besides, in the problem of quantum tunneling one of the $a$-coefficients in (8a) and (8b) can always be assumed to be known and real. We consider the coefficient $a_{r,2}$ to be such. Then the probability current continuity reads as

$$j_{l,12} a_{l,1} + j_{l,12}^* a_{l,1}^* = -j_{l,12} a_{r,1} - j_{l,12}^* a_{r,1}^* - 2j_{r,22} a_{r,2}. \tag{9}$$

This is a interrelation between the unknown coefficients $a_{l,1}$ and $a_{r,1}$. Thus, the discussion presented so far for unknown coefficients is summed up in two relations (6) and (9).

Let us to proceed to consideration of the quantum tunnelling problem with a wave falling on the barrier on the left side. To this end, we must refer to expressions of wave functions at asymptotically far distances $z \to -\infty$ and $z \to +\infty$. For the right hand side we get

$$\psi_r(z) \approx \frac{2^{-1+\frac{iu_0}{2\sqrt{\varepsilon}}}\left(i\sqrt{\varepsilon}\right)^{-1-\frac{iu_0}{2\sqrt{\varepsilon}}}}{\Gamma\left(1-\frac{iu_0}{2\sqrt{\varepsilon}}\right)} a_{r,1}\, z^{-\frac{iu_0}{2\sqrt{\varepsilon}}} e^{i\sqrt{\varepsilon}\,z} +$$

$$2^{-1+\frac{iu_0}{2\sqrt{\varepsilon}}}\left[\frac{\left(-i\sqrt{\varepsilon}\right)^{-1+\frac{iu_0}{2\sqrt{\varepsilon}}}}{\Gamma\left(1+\frac{iu_0}{2\sqrt{\varepsilon}}\right)} a_{r,1} + \left(i\sqrt{\varepsilon}\right)^{-1+\frac{iu_0}{2\sqrt{\varepsilon}}} a_{r,2}\right] z^{\frac{iu_0}{2\sqrt{\varepsilon}}} e^{-i\sqrt{\varepsilon}\,z},$$

where $\Gamma(\cdot)$ is the gamma function. The absence of a wave falling from the right side means the zero value of the second line of this equation, namely

$$a_{r,1} = -\frac{\left(i\sqrt{\varepsilon}\right)^{-1+\frac{iu_0}{2\sqrt{\varepsilon}}}}{\left(-i\sqrt{\varepsilon}\right)^{-1+\frac{iu_0}{2\sqrt{\varepsilon}}}} \Gamma\left(1+\frac{iu_0}{2\sqrt{\varepsilon}}\right) a_{r,2}, \tag{10}$$

which directly expresses the unknown coefficient $a_{r,1}$ through the chosen as known $a_{r,2}$.

The coefficient before the amplitude $a_{r,2}$ in (10) is a real quantity. Then the reality follows also for the amplitude $a_{r,1}$. In addition, since the coefficients $a_{l,1}$ and $a_{r,1}$ are at the regular parts of the general wave functions (5) and (3), the real character can be assumed also for the coefficient $a_{l,1}$ without any restriction on the singular character of the general solution. Then equation (9) reads as

$$a_{l,1} = -a_{r,1} - \frac{j_{r,22}}{\mathrm{Re}[j_{l,12}]} a_{r,2}, \tag{11}$$

which, taking into account (10), uniquely determines the coefficient $a_{l,1}$ through the $a_{r,2}$. As a result, equations (6), (10), and (11) define all three amplitudes in terms of $a_{r,2}$ and uniquely formulate the quantum tunneling problem. Crucially, they were obtained without addressing the question of the derivative of the wave function at the singularity point and therefore retained the singular character of the problem.

The asymptotic wave functions of incident, passing, and reflected waves have the following explicit expressions:

$$\psi_{incident} \approx -2^{-1+\frac{iu_0}{2\sqrt{\varepsilon}}}\left[\frac{\left(i\sqrt{\varepsilon}\right)^{-1+\frac{iu_0}{2\sqrt{\varepsilon}}}}{\Gamma\left(1+\frac{iu_0}{2\sqrt{\varepsilon}}\right)} a_{l,1} + e^{2i\pi\left(-1+\frac{iu_0}{2\sqrt{\varepsilon}}\right)}\left(-i\sqrt{\varepsilon}\right)^{-1+\frac{iu_0}{2\sqrt{\varepsilon}}} a_{l,2}\right] \left(\frac{1}{z}\right)^{-\frac{iu_0}{2\sqrt{\varepsilon}}} e^{i\sqrt{\varepsilon}\,z},$$

$$\psi_{transmitted} \approx \frac{2^{-1-\frac{iu_0}{2\sqrt{\varepsilon}}}\left(i\sqrt{\varepsilon}\right)^{-1-\frac{iu_0}{2\sqrt{\varepsilon}}}}{\Gamma\left(1-\frac{iu_0}{2\sqrt{\varepsilon}}\right)} a_{r,1}\, z^{-\frac{iu_0}{2\sqrt{\varepsilon}}} e^{i\sqrt{\varepsilon}\,z},$$

$$\psi_{reflected} \approx -\frac{2^{-1-\frac{iu_0}{2\sqrt{\varepsilon}}} e^{\frac{\pi u_0}{\sqrt{\varepsilon}}} \left(-i\sqrt{\varepsilon}\right)^{-1-\frac{iu_0}{2\sqrt{\varepsilon}}}}{\Gamma\left(1-\frac{iu_0}{2\sqrt{\varepsilon}}\right)} \cdot a_{1,1} \left(\frac{1}{z}\right)^{\frac{iu_0}{2\sqrt{\varepsilon}}} e^{-i\sqrt{\varepsilon}z}.$$

They define the corresponding probability currents according to definition (7) and the coefficients of transmission and reflection as their proper ratioes

$$T = \frac{j_{transmitted}}{j_{incident}}, \quad R = \frac{j_{reflected}}{j_{incident}} = 1 - T. \tag{12}$$

Their behaviour in the low energy range is shown in figure 1: as we approach the limit $\varepsilon = 0$, the oscillations thicken infinitely. This is the main reflection of singularity in the quantum tunnelling problem of a one-dimensional Coulomb potential barrier.

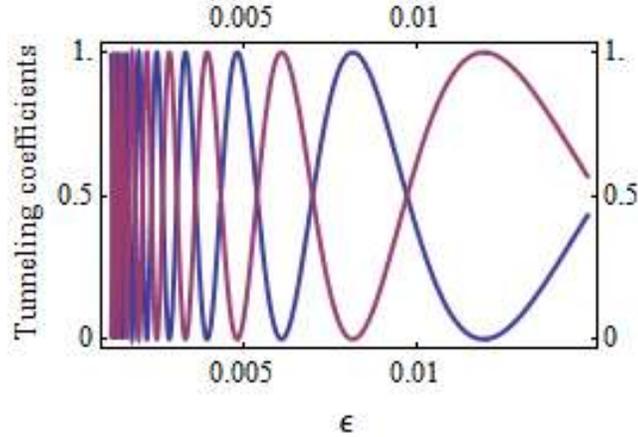

**Figure 1**. Transmission and reflection coefficients of a one-dimensional Coulomb potential barrier depending on the energy of the particle falling on it. The frequency of oscillations increases monotonously to infinity as the energy approaches the boundary $\varepsilon = 0$. The "power" of the potential barrier is $u_0 = 1$. Its decrease slows down the rate of oscillation.

When the energy of the particle moves to the high-energy region, the oscillation disappears approximately at the value $\varepsilon = u_0$ and the transmittance tends to unity, similar to the case of regular potentials. This transition is shown in figure 2.

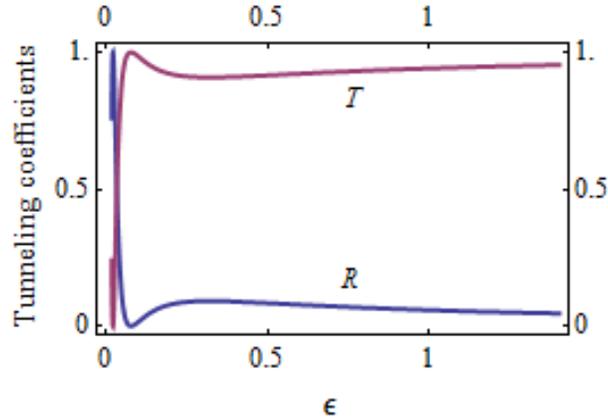

**Figure 2**. Continuation of the figure 1 to the region of higher energies. The oscillations are smoothed out, and further tunnelling of the singular potential behaves as in the case of a regular potential.

### 3. Conclusion

The problem of quantum tunnelling of a singular potential usually applies the regularization method, when the singularity is first removed in a narrow area around a singular point, the problem is solved for this prototype, and then in the coefficients of transmission and reflection, the limiting transition of narrowing the regularization region to zero is made. The other approach implies physically perceived conditions (Friedrichs extension in the Hilbert space) for matching the wave function and its derivative on both sides of the singularity point. The one-dimensional Coulomb potential in both approaches reveals itself as impenetrable. But with both approaches, the mathematical essence of the singularity is essentially suppressed, and so the question arises as to how the phenomenon of quantum tunnelling behaves while preserving the mathematical essence of the singularity. It is well known that in the case of a one-dimensional Coulomb potential, the singular character is significantly manifested in the derivative of the wave function, when it diverges logarithmically when approaching a singular point. Therefore, the problem of tunneling requires a formulation that does not address the question of the derivative of the wave function. This procedure is performed in this paper. It is found that the singularity gives the barrier permeability an oscillatory behavior, the frequency of which accelerates to infinity as the particle energy approaches the zero boundary. In addition, in the region of moderate energies, the oscillation disappears, the barrier tunnelling goes into the mode of regular potentials and becomes complete in the high-energy asymptote.

**Acknowledgements**
This work was supported by Armenian State Committee of Science.